\documentclass[pre, aps,amsmath,amssymb,preprint]{revtex4-1}
\usepackage{graphicx}
\usepackage{dcolumn}
\usepackage{bm}
\usepackage{color}
\usepackage{multirow}

\newcommand{\llangle}{\langle\!\langle}
\newcommand{\rrangle}{\rangle\!\rangle}
\renewcommand{\langle}{\left<}
\renewcommand{\rangle}{\right>}

\renewcommand{\hat}[1]{{\widehat #1}}
\newcommand{\comments}[1]{}

\begin{document}


\title{First passage time in multi-step stochastic processes with applications to dust charging}



\author{B. Shotorban}\email{ babak.shotorban@uah.edu}
\affiliation{Department of Mechanical and Aerospace Engineering, The University of Alabama in Huntsville, Huntsville, Alabama 35899}


\date{\today}

\begin{abstract}
An approach was developed to describe the first passage time (FPT) in multistep stochastic processes with discrete states governed by a master equation (ME). The approach is an extension of the totally absorbing boundary approach given for calculation of FPT in one-step processes (Van Kampen 2007) to include multistep processes where jumps are not restricted to adjacent sites.  In addition, a Fokker-Planck equation (FPE) was derived from the multistep ME, assuming the continuity of the state variable. The developed approach and an FPE based approach (Gardiner 2004) were used to find the mean first passage time (MFPT) of the transition between the negative and positive stable macrostates of dust grain charge when the charging process was bistable. The dust was in a plasma and charged by collecting  ions and electrons,  and emitting secondary electrons. The MFPTs for the transitioning of grain charge from one macrostate to the other were calculated by the two approaches for a range of grain sizes. Both approaches produced very similar results for the same grain except for when it was very small. The difference between MFPTs of two approaches for very small grains was attributed to the failure of the charge continuity assumption in the FPE description. For a given grain, the MFPT for a transition from the negative macrostate to the positive one was substantially larger than that for a transition in a reverse order. The  normalized MFPT for a transition from the positive to the negative macrostate showed little sensitivity to the grain radius. For a reverse transition, with the increase of the grain radius, it dropped first and then increased.  The probability density function of  FPT was substantially wider for a transition from the positive to the negative macrostate, as compared to a reverse transition.  

\end{abstract}

\pacs{}

\maketitle 


\section{Introduction}
\label{Introduction}
An intrinsic noise, characterized by random fluctuations, is inherent in the systems with particles  \cite{Kampen07}. An  electrically charged dust grain is an example of such systems where the charging is attributed to the ion and electron particles that are attached to or emitted from the grain. The electron and ion attachment  events occur at time intervals characterized by randomness. Hence, the net charge of the grain  exhibits random fluctuations over time. Intrinsic noises can be described by a master equation (ME) governing the probability density function (PDF) of the state of the system~\cite{Kampen07}. 
The ME of the dust charging system admits different forms where the functionality of the transition probability rates in terms of the charge state depends on the mechanisms that are responsible for charging. These forms are previously constructed when the charging mechanisms are the collisional collection of electrons and singly charged positive ions \cite{MR95,MRS96,MR97,shotorban2011nonstationary}, the collisional collection of electrons and multiply charged ions \cite{shotorban2014intrinsic,mishra2015statistical}, and the collisional collection of ions and electrons combined with the secondary emission of electrons (SEE) from the grain \cite{gordiets1998charge,gordiets1999charge,shotorban2015bistable}.

Intrinsic charge fluctuations of a grain were the subject of several investigations \cite{morfill1980dust,CG94,MR95,MRS96,MR97,gordiets1998charge,KNPV99,gordiets1999charge,shotorban2011nonstationary,asgari2011stochastic,shotorban2012stochastic,matthews2013cosmic,shotorban2014intrinsic,asgari2014non,mishra2015statistical,shotorban2015bistable,matthews2018discrete}.  
Let $Z(t)$ indicate the instantaneous net elementary charge (charge state) of the  grain. It is an integer variable and a function of time $t$. When, for example, five singly charged positive ions and eight electrons are collected on the grain at a given time, the grain is at charge state $Z=+5-8=-3$. Here, $Z(t)$ experiences stepwise variation over time, because it can only have integer values, as a consequence of discreteness nature of  charge. \citet{morfill1980dust} suggested that intrinsic fluctuations of $Z(t)$ follows  $Z_\mathrm{rms}^2\propto |\langle Z \rangle|$, where $\langle Z \rangle$ and $Z_\mathrm{rms}^2$ denote the mean and variance of $Z(t)$, respectively.  At the limit of large $|Z(t)|$, the charge is assumed to continuously vary and accordingly, the grain charge probability distribution is shown to be Gaussian with the mean and variance found to respect the proportionality correlation above \cite{MR95,shotorban2011nonstationary,shotorban2014intrinsic}. This Gaussianity was determined  for situations where the mechanism of charging was the collisional collection of plasma particles. A similar finding was also made for situations where  the  thermionic emission or UV irradiation mechanisms are active dust charging mechanisms in addition to the collisional collection  mechanisim \cite{KNPV99}. On the other hand, it was shown that when $|Z(t)|$ is smaller than tens of elementary charges, the charge probability distribution can significantly deviate from  Gaussianity \cite{MR95,shotorban2011nonstationary}. This deviation is attributed to the significance of charge discreetness effects in the time evolution of $|Z(t)|$ when it is small. The smaller $|Z(t)|$ is, the larger the deviation is. In the studies reviewed above, the dust charge fluctuations were stable, a feature characterized by fluctuations around a fixed stable point (stationary stable macrostate).

It was shown \cite{shotorban2015bistable} that if both collisional collection and SEE mechanisms were active, charge fluctuations could be unstable. This instability  was characterized by a substantial deviation of the grain charge PDF  from Gaussianity in some cases.    Moreover, it was shown that if the SEE was active, fluctuations could be bistable. This  bistability is associated with a bifurcation phenomenon of the grain charge,  known from the investigations of \citet{meyer1982flip} and \citet{horanyi1998electrostatic} who accounted for SEE while neglecting charge fluctuations. In this phenomenon, two identical grains in the same plasma environment exhibit two distinct (mean) charge values, one positive and the other negative. \citet{shotorban2015bistable} showed that these  values correspond to two macrostates \cite{Kampen07} between which the charge intrinsic fluctuations of a grain can switch.  This behavior is known as metastability in stochastic processes and it is a state where  the fluctuations are characterized by two distinct time scales - one associated with the fluctuations at either macrostate and the other  with the spontaneous transition between the mactrostates. It was shown  that a  switch from the negative macrostate to the positive macrostate is attributed to a sequence of incidents with ion attachment or  primary electron attachment that resulted in emission of secondary electrons \cite{shotorban2015bistable}.  On the other hand, a reverse switch  is attributed to a sequence of incidents most of which are the attachments of primary electrons that result in no emission of secondary electrons.

The current study was motivated by a need to determine the first-passage time (FPT) \cite{Kampen07} of  grain charge fluctuations: Starting from a given charge, how long does it take for the grain to posses a specified charge? The FPT is of particular interest in metastable fluctuations, as it quantifies the time scales of the transitions between the macrostates. The FPT is a random quantity whose behavior can be described by the calculation of its statistical properties.  For example, the growth and dissipation times, calculated by \citet{MR97} for grain charging due to collisional collection of plasma particles, are special cases of the mean first passage time (MFPT).  The growth time is defined as the mean time for the transition from the mean charge to a specified charge, and the dissipation time is the mean time to revert from the specified charge to the mean charge.  
Recently, \citet{matthews2018discrete} used \citet{MR97}'s formulation for the growth and dissipation times, to validate a discrete stochastic model for the charging of aggregate grains.  This formulation is limited to the grain charging described by a linear Fokker-Planck equation (FPE) \cite{Kampen07} derived from  the ME of a one-step process. 
In the next section, mathematical approaches are proposed to calculate the FPT of multi-step processes. Then, they are used to investigate the FPTs in the grain charging system with a focus on bistable situations. 

\section{First Passage Time in Multistep Processes}

Consider  a  stochastic process with a discrete set of states governed by the following master equation:  
\begin{eqnarray}
\frac{dP(Z)}{dt}&=&\sum_{n=1}^N\left[r_n(Z+n)P(Z+n)-r_n(Z)P(Z)\right]\nonumber\\
&+&\sum_{n=1}^M\left[g_n(Z-n)P(Z-n)-g_n(Z)P(Z)\right],
\label{eq:masterEquation}
\end{eqnarray}
where $Z$ is an integer indicating the state variable (site), e.g., the net elementary charge possessed by a grain, and $P(Z)$ is the probability density function. Here,  $r_n(Z)$ is the probability per unit time that, being at $Z$ a jump occurs to $Z-n$ and $g_n(Z)$ is the probability per unit time that, being at $Z$ a jump occurs to $Z+n$.   The process modeled by eq. (\ref{eq:masterEquation}) can be regarded as a ``multistep process'', a generalized notion of the one-step process \cite{Kampen07}, where jumps can also occur between non-adjacent sites. The change of the grain charge by multiple units as a result of collecting multiply charged  ions \cite{mamun2003charging} is an example of a jump between non-adjacent sites \cite{gordiets1999charge,shotorban2014intrinsic,mishra2015statistical}. The other example is when multiple secondary electrons are emitted when a primary electron impacts the grain \cite{gordiets1999charge,shotorban2014intrinsic}.   The one-step process master equation is a special case of eq. (\ref{eq:masterEquation}) with $N=M=1$. The master equations formulated for grain charging in multicomponent plasma \cite{shotorban2014intrinsic} and in cases where the SEE is active \cite{shotorban2015bistable}, can be readily recast in the form given in eq.~(\ref{eq:masterEquation}), as illustrated in \S~\ref{sec:stochasticCharging}.

The notion of the macroscopic or phenomenological equation illustrated by \citet{Kampen07} for one-step processes is extended here to include multistep processes.  That is a deterministic differential equation where the fluctuations of $Z(t)$ is ignored and treated as a non-stochastic quantity. An approach to obtain this equation is to multiply master eq.~(\ref{eq:masterEquation}) by $Z$ and sum over $Z$:

\begin{equation}
\frac{d\langle Z\rangle}{dt}=-\sum_{n=1}^N n \langle r_n(Z)\rangle+\sum_{n=1}^M n \langle g_n(Z)\rangle,
\label{eq:masterEquationAve}
\end{equation}
where $\langle\rangle$ indicates the mean defined for a function such as $\alpha(Z)$ by $\langle\alpha(Z)\rangle=\sum_Z\alpha(Z)P(Z)$. In the derivation of eq.~(\ref{eq:masterEquationAve}), the index shift identity of the summation manipulation  role is used and $P(Z)=0$ is assumed at the boundaries.  Unless $r_n(Z)$ and $g_n(Z)$ are linear functions of $Z$, eq.~(\ref{eq:masterEquationAve}) is not a closed equation. However, if they are nonlinear, they may be expanded about $\langle Z\rangle$, e.g.,  
\begin{equation}
    \langle r_n(Z)\rangle= r_n(\langle Z\rangle )+\frac{1}{2}\langle(Z-\langle Z\rangle)^2\rangle r_n''(\langle Z\rangle)+\dotsb.
\label{eq:expansion}
\end{equation}
This expansion shows that higher-order moments play a role in the time evolution of $\langle Z\rangle$ through (\ref{eq:masterEquationAve}). Nonetheless, 
retaining only the first term in the expansion above and the one for $\langle g_n(Z)\rangle$, the macroscopic equation is obtained

\begin{equation}
\frac{d\cal Z}{dt}=-\sum_{n=1}^N n  r_n({\cal Z})+\sum_{n=1}^M n  g_n({\cal Z}),
\label{eq:macroscopic}
\end{equation}
where ${\cal  Z}\equiv\langle Z\rangle$. 

\subsection{First Passage Time in Master Equation}
Here, to calculate the FPT, the absorbing boundary approach available for  one-step processes \cite{Kampen07}, is extended  to include multistep processes:

Suppose that the system is at state $Z=m$ at $t=0$. To calculate the escape time to a state $Z\ge R$, where $R$ is located on the right of $m$, i.e., $m<R$, eq.~(\ref{eq:masterEquation}) is solved in the range $Z<R$ with the initial condition $P(Z)=\delta_{Z,m}$ (Kronecker delta function). Here, $R$ is set as the {\em totally absorbing} boundary condition by setting 
$P(Z)=0$ if $Z\ge R$ for all  times. 

The probability for the system to be at a site in the domain $-\infty < Z<R$ is $\sum_{Z=-\infty}^{R-1}P(Z)$. Now let $f_{R,m}(t)dt$ indicate the probability  that starting at site $m$, the system reaches $R$ or beyond at a time between $t$ and $t+dt$. Then:

\begin{eqnarray}
f_{R,m}(t)&=&-\frac{d}{dt}\sum_{Z=-\infty}^{R-1}P(Z,t)\nonumber\\
&=&\sum_{n=1}^M\sum_{Z=R-n}^{R-1}g_n(Z)P(Z,t),
\label{eq:fRm}
\end{eqnarray}

\noindent
where the second line is derived from  the substitution for $dP(Z)/ dt$ in the first,  using the derivative of eq.~(\ref{eq:masterEquation}) and the  manipulation below:

\begin{widetext}
\begin{eqnarray}
\sum_{Z=-\infty}^{R-1}\frac{dP(Z)}{dt}&=&\sum_{n=1}^N\sum_{Z=-\infty}^{R-1}\left[r_n(Z+n)P(Z+n)-r_n(Z)P(Z)\right] \nonumber\\
&&+\sum_{n=1}^M\sum_{Z=-\infty}^{R-1}\left[g_n(Z-n)P(Z-n)-g_n(Z)P(Z)\right] \nonumber\\
&=&\sum_{n=1}^N\left[\sum_{Z=-\infty}^{R-1+n}r_n(Z)P(Z)-\sum_{Z=-\infty}^{R-1}r_n(Z)P(Z)\right]\nonumber\\
&&+\sum_{n=1}^M\left[\sum_{Z=-\infty}^{R-1-n}g_n(Z)P(Z)-\sum_{Z=-\infty}^{R-1}g_n(Z)P(Z)\right]\nonumber\\
&=&\sum_{n=1}^N\sum_{Z=R}^{R-1+n}r_n(Z)P(Z)-\sum_{n=1}^M\sum_{Z=R-n}^{R-1}g_n(Z)P(Z)\nonumber\\
&=&-\sum_{n=1}^M\sum_{Z=R-n}^{R-1}g_n(Z)P(Z)
\end{eqnarray}
\end{widetext}
 noting that in the line before the last one, the first term vanishes since $P(Z)=0$ for $Z\ge R$.

 On the other hand, the total  probability of reaching a state at $Z\ge R$ is calculated by

\begin{eqnarray}
\pi_{R,m}&=&\int_0^\infty f_{R,m}(t)dt = 1-\sum_{Z=-\infty}^{R-1} P(Z,t=\infty)\nonumber\\
&=&\sum_{n=1}^M\sum_{Z=R-n}^{R-1}g_n(Z)\int_0^\infty P(Z,t)dt,
\end{eqnarray}
 and the MFPT is
\begin{eqnarray}
\tau_{R,m}&=&\frac{1}{\pi_{R,m}}\int_0^\infty t f_{R,m}(t)dt \nonumber \\
&= &\frac{1}{\pi_{R,m}}\sum_{n=1}^M\sum_{Z=R-n}^{R-1}g_n(Z)\int_0^\infty t P(Z,t)dt.
\end{eqnarray}

 Likewise, for the calculation of the FPT from  the  state $m$ to the state $Z\le L$ where $L<m$, eq.~(\ref{eq:masterEquation}) is solved for $P(Z)$ in the domain $L<Z$ with $L$  set as the totally absorbing boundary condition, i.e., $P(Z)=0$ if $Z\le L$. Then
 \begin{equation}
f_{L,m}(t)=-\frac{d}{dt}\sum_{Z=L+1}^\infty P(Z,t)=\sum_{n=1}^N\sum_{Z=L+1}^{L+n}r_n(Z)P(Z,t).
\label{eq:fLm}
\end{equation}
The total  probability of reaching $L$ or beyond is

\begin{eqnarray}
\pi_{L,m}&=&\int_0^\infty f_{L,m}(t)dt = 1-\sum_{Z=L+1}^{\infty} P(Z,t=\infty)\nonumber\\
&=&\sum_{n=1}^N\sum_{Z=L+1}^{L+n}r_n(Z)\int_0^\infty P(Z,t)dt,
\end{eqnarray}
 and the MFPT is

\begin{eqnarray}
\tau_{L,m}&=&\frac{1}{\pi_{L,m}}\int_0^\infty t f_{L,m}(t)dt \nonumber \\
&= &\frac{1}{\pi_{L,m}}\sum_{n=1}^N\sum_{Z=L+1}^{L+n}r_n(Z)\int_0^\infty t P(Z,t)dt.
\end{eqnarray}

\subsection{Mean First Passage Time in Fokker-Planck Equation}

If $r_n(Z)$ and $g_n(Z)$ are smooth functions of $Z$, i.e. continuous and differentiable   a number of times, $g_n(Z)$ slightly change between $Z$ and $Z+M$, and $r_n(Z)$ slightly between $Z$ and $Z+N$, $Z$ may be treated as a continuous variable. Hence, expanding the first terms in the summations in eq.~(\ref{eq:masterEquation}) by Taylor's series  and retaining the terms up to the second derivative, e.g.,

\begin{eqnarray}
r_n(Z\pm n)P(Z\pm n)&=&r_n(Z)P(Z)\pm n\frac{\partial}{\partial Z} \left[ r_n(Z)P(Z)\right]\nonumber\\
&+&\frac{n^2}{2}\frac{\partial^2}{\partial Z^2}\left[ r_n(Z)P(Z)\right],
\label{eq:Taylor}
\end{eqnarray}
\noindent
a (forward) Fokker-Planck equation can be derived: 
\begin{equation}
\frac{\partial P(Z,t)}{\partial t}=-\frac{\partial}{\partial Z} A(Z)P(Z)\\
+\frac{1}{2}\frac{\partial^2}{\partial Z^2}B(Z)P(Z),
\label{eq:FPEquation}
\end{equation}
where the drift and diffusion functions are 
\begin{equation}
    A(Z)=-\sum_{n=1}^N n r_n(Z)+\sum_{n=1}^M n g_n(Z),
\label{eq:Drift}
\end{equation}
\begin{equation}
    B(Z)=\sum_{n=1}^N n^2 r_n(Z)+\sum_{n=1}^M n^2 g_n(Z),
\label{eq:Diffusion}
\end{equation}
respectively. Eq.~(\ref{eq:FPEquation}) is identical to the Fokker-Planck eq.~(\ref{eq:FPAppendix})  in Appendix \ref{sec:FirstPassageTime}, where the formulation provided by \citet{Gardiner04}  for the calculation of MFPT in the FPE, is illustrated. It is noted that  the drift $A(Z)$, given in eq.~(\ref{eq:Drift}), is identical to the r.h.s. of eq.~(\ref{eq:macroscopic}). On the other hand, since $r_n$ and $g_n$ are positive functions, the positivity of the diffusion coefficient $B(Z)$ is secured  in eq.~(\ref{eq:Diffusion}).  

The macroscopic equation associated with eq.~(\ref{eq:FPEquation}) is obtained by  integrating it after multiplication by $Z$,  using an expansion similar to eq.~(\ref{eq:expansion}) and  retaining the lowest order term. The resulting macroscopic equation is  identical to  eq.~(\ref{eq:macroscopic}), which is the macroscopic equation associated with the master eq.~(\ref{eq:masterEquation}). The stationary macrostates of the system is defined by the roots of  the r.h.s. of eq.~(\ref{eq:macroscopic}), viz. $A({\cal Z}^s)=0$. If the condition $A'({\cal Z}^s)<0$ is satisfied, the associated macrostate is stable. A detail discussion on stability, instability and bistablity of stochastic processes can be found in Ref. \onlinecite{Kampen07}. 

In general, function $A(Z)$ is nonlinear. However, if $A(Z)=0$ has only one stable  root $Z={\cal Z}^s$ (stable stationary macrostate), or there are more but at least one root is sufficiently away from the rest, then $A(Z)$ can be linearized about this macrostate.  Consequently, a linear FPE, where drift is linear and  diffusion coefficient is constant \cite{Kampen07}, can be derived. This derivation is achieved by expanding  $A(Z)$ and $B(Z)$ about ${\cal Z}^s$, retaining the lowest non-zero term, and substituting them in eq.~(\ref{eq:FPEquation}):

\begin{eqnarray}
\frac{\partial P(Z,t)}{\partial t}&=&- A'\left({\cal Z}^s\right)\frac{\partial}{\partial Z}\left(Z-{\cal Z}^s\right)P(Z)\nonumber\\
& &+\frac{1}{2}B\left({\cal Z}^s\right)\frac{\partial^2}{\partial Z^2}P(Z), 
\label{eq:FPEquationLinear}
\end{eqnarray}
where using eq.~(\ref{eq:Drift}),
\begin{equation}
  A'(Z)=\frac{dA}{dZ}=  -\sum_{n=1}^N n r_n'(Z)+\sum_{n=1}^M n g_n'(Z).
\end{equation}
Eq.~(\ref{eq:FPEquationLinear}) is a linear FPE, which is valid for fluctuations at the vicinity of ${\cal Z}^s$.  This equation describes the Ornstein-Uhlenbeck process with a Gaussian solution at a stationary state with a mean and variance of 
\begin{equation}
    \langle Z\rangle^s={\cal Z}^s,
\end{equation}
\begin{equation}
    \llangle Z\rrangle^s=-\frac{1}{2}\frac{B({\cal Z}^s)}{A'({\cal Z}^s)},
\end{equation}
respectively, and a correlation of
\begin{equation}
    \llangle Z(t)Z(t+u)\rrangle^s=\llangle Z\rrangle^s\exp\left(-\frac{u}{\tau_0}\right),
\end{equation}
where $\tau_0=-1/A'({\cal Z}^s)$.  

The procedure outlined in Appendix \ref{sec:FirstPassageTime} can be also used to calculate MFPT in the linear FPE. Two specific MFPTs are the growth and dissipation times defined  in \S\ref{Introduction}.  The growth time is calculated by setting $L={\cal Z}^s-Z$, $R={\cal Z}^s+Z$ and $y={\cal Z}^s$ in the integral solution  in eq.~(\ref{eq:leftOrRight}), which is simplified for the linear FPE to   

\begin{equation}
    \frac{\tau_Z\left({\cal Z}^s\right)}{\tau_0}=
    \frac{\Delta_Z^2}{2}\, _2F_2\left(1,1;\frac{3}{2},2;
     \frac{\Delta_Z^2}{2}\right),
\label{eq:growthTime}
\end{equation}
where $\Delta_Z=|Z-{\cal Z}^s|/
    \sqrt{\llangle Z\rrangle^s}$ and $_2F_2$ is a generalized hypergeometric function \cite{Weisstein}. On the other hand, the dissipation time is calculated by eq.~(\ref{eq:rightToLeft})

\begin{equation}
    \frac{\tau_s\left(Z\right)}{\tau_0}=\frac{1}{2}\pi\,  
    \mathrm{erfi}\left(\frac{\Delta_Z }{\sqrt{2}}\right)
    - \frac{\tau_Z\left({\cal Z}^s\right)}{\tau_0}.   
\label{eq:dissipationTime}
\end{equation}
Eqs.~(\ref{eq:growthTime}) and (\ref{eq:dissipationTime}) are in a simplified form of the integral solutions previously provided \cite{Kampen07,MR97} for a linear FPE.  It is noted that, here, the  process is not restricted to the one step assumption previously made~\cite{MR97}.

\begin{figure}
\begin{center}
\includegraphics[width=0.5\columnwidth, angle=0]{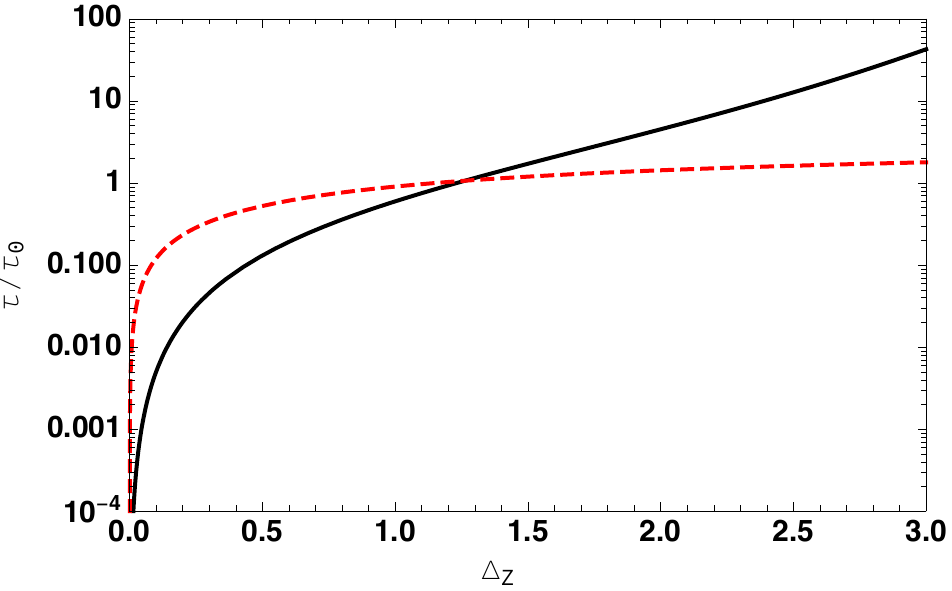}
\end{center}
\caption{Dimensionless growth (solid line) and dissipation (dashed line)  times vs dimensionless deviation of the state variable from its mean when the PDF is governed by a linear FPE.}
\label{fig:GrowthDissipation}
\end{figure}

Figure \ref{fig:GrowthDissipation} displays the dimensionless growth time $\tau_Z({\cal Z}^s)/\tau_0$ and the dimensionless dissipation time $\tau_s(Z)/\tau_0$ versus  $\Delta_Z$. Both times increase monotonically from zero. The dissipation time experiences a steep increase initially but its rate of increase rapidly drops. The growth time starts off with a slower rate but intersects with the dissipation time at $\Delta_Z =1.26278$ and ${\tau_Z\left({\cal Z}^s\right)}/{\tau_0}={\tau_s\left(Z\right)}/{\tau_0}=1.06319$. Then, the difference between the growth and dissipation times grows, becoming an order of magnitude larger at $\Delta_Z \sim 3$.  

\section{Stochastic Charging of a grain in a Plasma}
\label{sec:stochasticCharging}
Consider a plasma with ion (electron) density of $n_{i(e)}$, temperature of $T_{i(e)}$, and mass of $m_{i(e)}$, and let   $\lambda_{D}=\sqrt{\epsilon_0k_BT_e/n_ee^2}$ and $\omega_{pe}=\sqrt{n_e e^2/\epsilon_0m_e}$ represent the Debye length and plasma frequency, respectively.  Moreover, let $I_\mathrm{i}(Z)$, $I_\mathrm{e}(Z)$ and $I_\mathrm{s}(Z)$ indicate the currents of ions, primary electrons and secondary emitted electrons from the grain, respectively. Ions are assumed singly positively charged, however, the discussion here can be extended to include multiply charged ions \cite{shotorban2014intrinsic}.  Let $f_j(Z)$ represent the probability distribution of $j$ electrons emitted from the grain in a single incident of primary electron attachment. This quantity is equivalent to the fraction of primary electrons that cause $j$ secondary electrons to emit in a single primary electron attachment incident. If $K$ represents the maximum number of secondary electrons that can be emitted in a single incident of the electron attachment, then $0\le j \le K $ and $\sum_{j=0}^Kf_j(Z)=1$.  It can be shown that $I_s(Z)=I_e(Z)\sum_{j=1}^Kf_j(Z)$.    If $N=1$, $r_1(Z)=f_0(Z)I_e(Z)$, $M=K-1$, and  $g_n(Z)=f_{n+1}(Z)I_e(Z)+\delta_{1n}I_i(Z)$ in eq.~(\ref{eq:masterEquation}), where $\delta_{mn}$ is the Kronecker delta function, the master equation governing grain charging with ions, electrons and SEE in a plasma ~\cite{shotorban2015bistable}. Using the rate equations above, the drift coefficient in eq.~(\ref{eq:FPEquation}) is simplified to $A(Z)=-I_e(Z)+I_i(Z)+I_s(Z)$ and the r.h.s. of the macroscopic eq.~(\ref{eq:macroscopic}) is the net current to the grain. 

The calculation of the currents of ions, primary electrons and secondary electrons and the probability distribution of emission of secondary electrons is illustrated in Appendix \ref{sec:Currents}  with the significant parameters noted below.  A reference grain charge is defined by 
\begin{equation}
\Omega=\frac{4 \pi \epsilon_0ak_BT_e}{e^2 },
\label{eq:omega}
\end{equation}
where $a$ represents the radius of the grain. It is noted  that $\Omega$ defined by eq.~(\ref{eq:omega}) was used as the system size \cite{Kampen07} in the stochastic description of grain charging  \cite{shotorban2011nonstationary,shotorban2014intrinsic,shotorban2015bistable}. Also, a reference charging time scale is defined by
\begin{equation}
\tau_c =  \frac{\Omega}{\Gamma}=\frac{\sqrt{2\pi}\lambda_{D}}
{\omega_\mathrm{pe}a}, 
\end{equation}
where 
\begin{equation}
\Gamma=\pi a^2n_e  \sqrt{8k_BT_e\over\pi m_e } =\frac{\Omega \omega_\mathrm{pe}a}{\sqrt{2\pi}\lambda_{D}},
\label{eq:Gamma}
\end{equation}
which indicates the electron current to the uncharged grain. For SEE from the grain, $T_s$ represents the temperature of the emitted secondary electrons, $\delta_M$ is the maximum yield which is around unity for metals and at the order 2 to  30 for insulators, and $E_M$ is the peak primary electron energy, a model constant ranging from 300 to 2000~eV. The values of these two parameters for various dust materials can be found in Ref.~\onlinecite{meyer1982flip}.


\begin{figure}
\begin{center}
\includegraphics[width=0.5\columnwidth, angle=0]{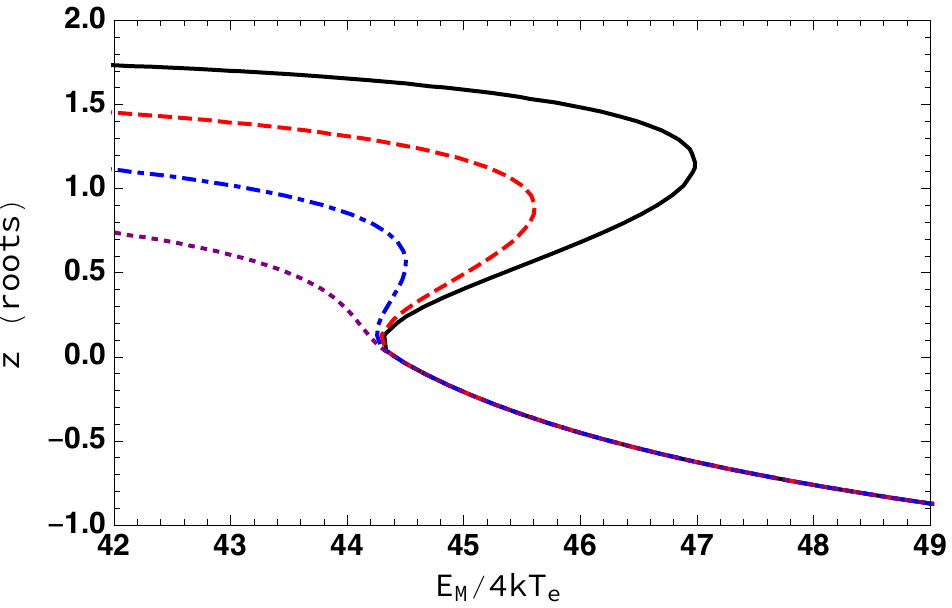}
\end{center}
\caption{Stationary macrostates (roots of the macroscopic equation at a stationary state) against $E_M/4kT_e$ near a triple-root situation for $E_M/4kT_s=30$ (solid line), $32$ (dashed line),  $35$ (dot-dashed line), and $40$ (dotted line);  $\delta_M=15$.}
\label{fig:S-Shape}
\end{figure}

The stationary grain charge macrostates, which correspond to the roots of eq. (\ref{eq:macroscopic}) with the l.h.s set to zero,  are plotted against $E_M/4kT_e$ in fig.~\ref{fig:S-Shape}. Here, $z=Z/\Omega$ indicates the normalized charge. The four curves correspond to four different values of $E_M/4kT_s$. It could be seen in this figure that they collapse into a single curve for $z<0$, which is attributed to the SEE current being independent from the SEE temperature as evident in eq.~(\ref{eq:SEE}) for  $z<0$. A triple root situation is observed for  $E_M/4kT_s=$30, 32 and 35 while this situation does not encounter for $E_M/4kT_s=40$. In the triple root situations, one of the roots is negative and stable (negative charge state) and the other two positive. The larger positive root is stable (positive charge state)  and  the smaller one is unstable.  A single positive root situation is encountered on the left of the triple root region while a single negative root situation is on the right  of this region.

\section{Results and Discussion}

Grains with a radius in the range of $10\sim100$nm suspended in a hydrogen plasma with $n_{i(e)}=10^4$~$\mathrm{m}^{-3}$, $T_i=T_e=2\times10^4$~K was considered. These values are relevant to interstellar dusty plasma condition \cite{Kimura1998}.

\begin{figure}
\begin{center}
\includegraphics[width=0.5\columnwidth, angle=0]{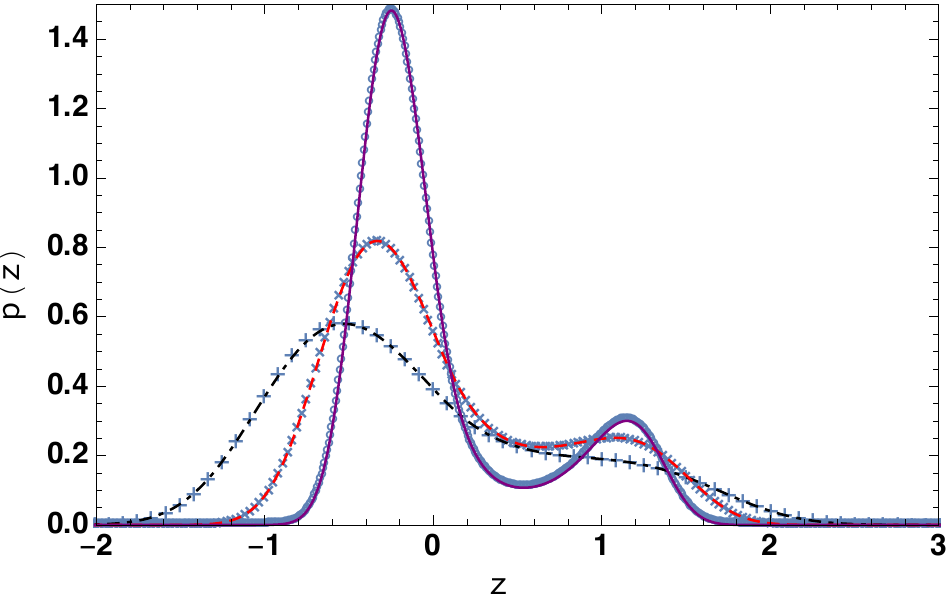}
\end{center}
\caption{Probability density function of the normalized grain charge ($z=Z/\Omega$) at a bistable state for grain radius of 100, 30 and 10nm shown by solid, dashed and dotted-dashed lines, respectively, through the Fokker-Planck equation, and shown by $\circ$, $\times$ and $+$, respectively, through the master equation; $\delta_M=15$, $E_M/4kT_e=45$, $E_M/4kT_s=32$.}
\label{fig:MaxPDF}
\end{figure}

Figure \ref{fig:MaxPDF} shows the PDF of the normalized charge obtained by solving the master equation and separately by solving the Fokker-Planck equation, for a bistable state of grain charging (a triple root situation where there are two stable stationary macrostates) for three different grain sizes. The agreement between ME and FPE solutions is excellent. A  bimodal distribution  is distinguishable for a grain radius of $100$nm. The distribution of the grain charge exhibits bimodality however it is less obvious for a grain radius of $30$nm.  

\begin{figure}
\begin{center}
\includegraphics[width=0.5\columnwidth, angle=0]{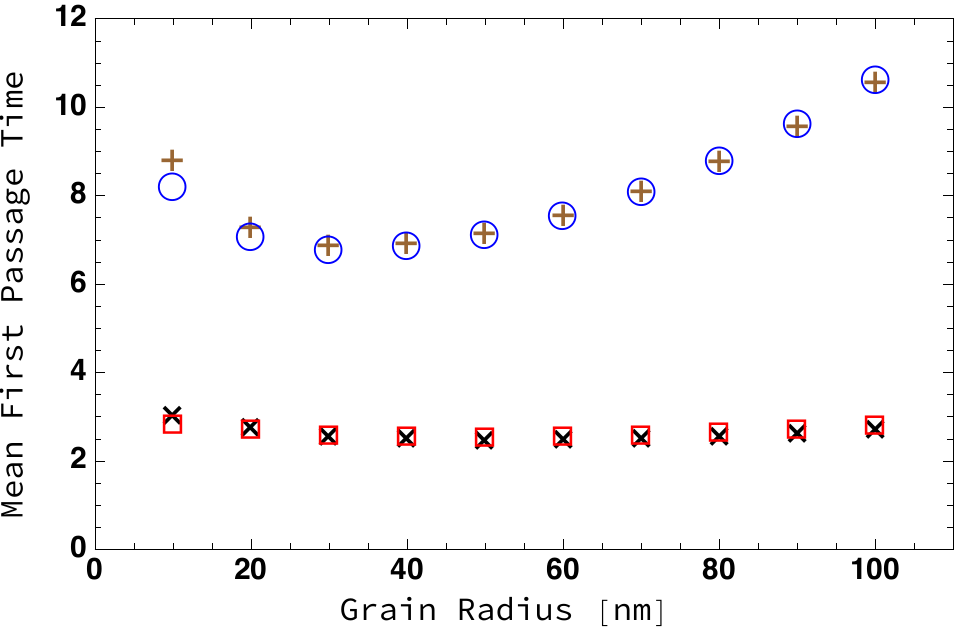}
\end{center}
\caption{Dimensionless MFPT versus grain radius for a transition from the negative macrostate to the positive macrostate through master equation ($+$) and Fokker-Planck equation ($\bigcirc$) and for a transition from the positive macrostate to the negative macrostate through master equation ($\times$) and Fokker-Planck equation ($\Box$); see caption of fig. \ref{fig:MaxPDF} for parameter values.}
\label{fig:meanFPT}
\end{figure}

Figure~\ref{fig:meanFPT} displays MFPT normalized by $\tau_c$, versus  grain radius. MFPT is calculated, using ME, and separately using the FPE. In the former approach, MFPT is obtained by integrating the PDF of FPT given in eqs.~(\ref{eq:fRm}) and (\ref{eq:fLm}). In the latter approach, it is obtained by eqs.~(\ref{eq:tauRight}) and (\ref{eq:rightToLeft}). Except for $R<30$nm, excellent agreement is seen between  two approaches in fig~\ref{fig:meanFPT}. For $R<30$nm, the difference between the approaches is much more pronounced for the dimensionless MFPT of a transition from the negative macrostate  to the positive state, compared to the one from the positive macrostate to the negative macrostate. The significant  difference between ME and FPE results here could be attributed to the discreteness of charge, which is neglected in FPE but it is more critical for smaller grains.  
For the grain radius range considered here, the dimensionless MFPT of a transition from the stable positive charge macrostate  to the negative charge  macrostate experiences little change, exhibiting a constant value of around 2.5.   On the other hand, the one from the negative charge macrostate  to the stable positive charge macrostate first descends,  reaching a minimum value of around seven at a grain radius of around $30$nm, and then gradually rises with the increase of radius. The radius at which this minimum occurs seems to be correlated with how the shape of PDF of the grain charge changes with radius (fig.~\ref{fig:MaxPDF}). As seen in fig.~\ref{fig:MaxPDF},  there is a deep saddle point for the PDF for a radius of 100nm whereas it does not exist for a radius of 10nm. A saddle point is also seen for a radius of 30nm; however, it is very shallow. It is found that the saddle point is deeper for a larger radius. At a given radius the MFPT from the negative macrostate to the positive macrostate is substantially larger than that from the positive macrostate  to the negative macrostate, indicating that the system fluctuates longer at the negative charge state.  

\begin{figure}
\begin{center}
\includegraphics[width=0.5\columnwidth, angle=0]{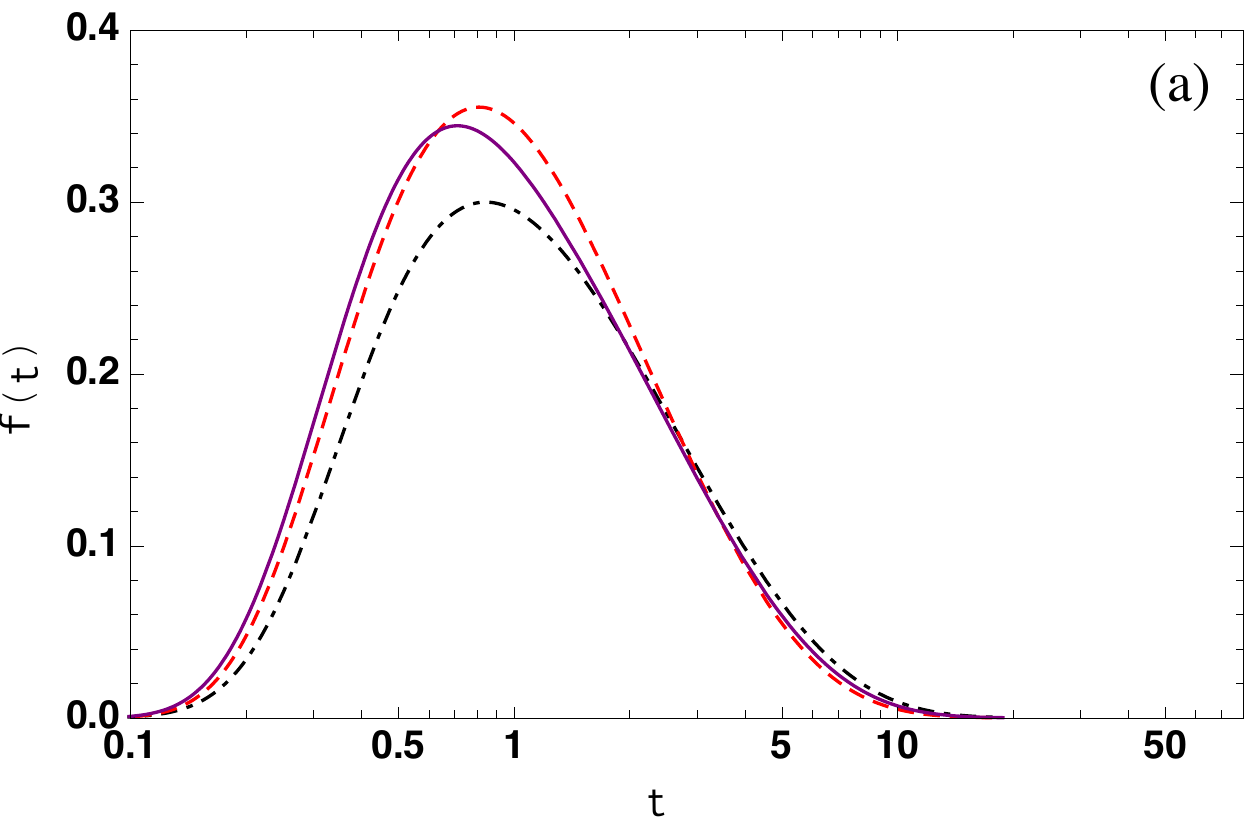}
\end{center}
\begin{center}
\includegraphics[width=0.5\columnwidth, angle=0]{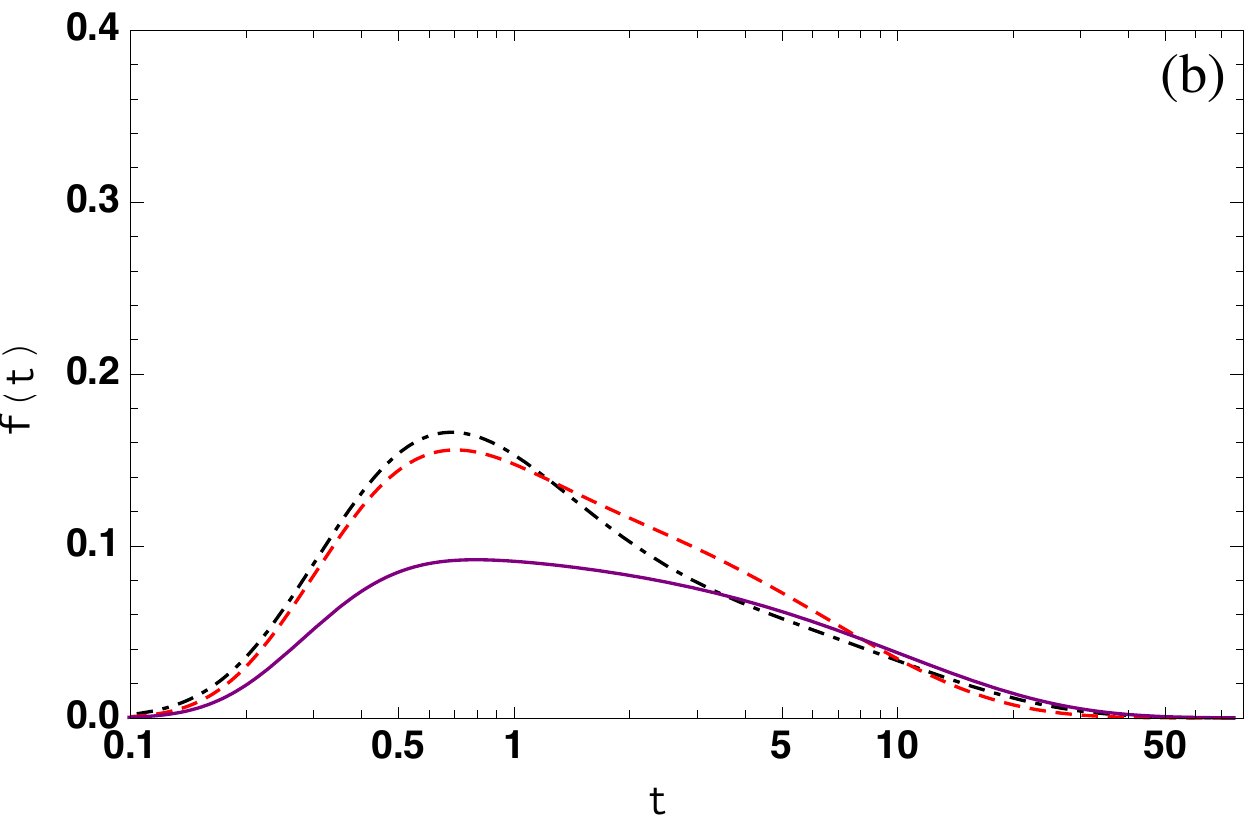}
\end{center}
\caption{Probability density function of the dimensionless FPT for grain radius of 100 (solid line), 30 (dashed line) and 10 nm (dotted-dashed line) for grain charge transitioning (a) from the positive macrostate to the negative macrostate; and (b) from the negative macrostate  to the positive  macrostate. See the caption of fig. \ref{fig:MaxPDF} for parameter values.}
\label{fig:FPTdistribution}
\end{figure}

Figure \ref{fig:FPTdistribution} displays the PDF  of the normalized FPT for three grain radii of 10, 30 and 100nm, which are calculated, using eqs.~(\ref{eq:fRm}) and (\ref{eq:fLm}). The PDFs in the bottom panel, which is for the transition from the stable positive to negative  macrosstate,  are wider than those in the top panel, which is for the reverse transition. 
In the top panel, the PDFs are similar for a dimensionless FPT larger than around 2. The grain charge is more populated around the negative charge macrostate for all three grain charge sizes, compared to the stable positive macrostate. That means a grain charge starting from the left macrostate  will overall remain longer in this macrostate before transiting to the postive macrostate.  

\section{Summary and Conclusions}
\label{sec:conclusions}
The  absorbing boundary  approach previously developed for  calculation of  FPT  in the stochastic processes that are governed by one-step master equations \cite{Kampen07}, was extended to include multi-step master equations. The restriction of jumps between adjacent sites in one step process is relaxed in multistep processes. The outcome of this extension was formulas for calculation of MFPT and the PDF of FPT.  The new approach was used to study FPT in the grain charging system where a grain is charged by collecting ions and electrons from a  plasma, and emitting secondary electrons as a result of the impact of the primary electrons. Depending on the plasma and grain parameters, such a grain charge system could have only one stationary stable macrostate or two stationary stable  macrostates (bistable), one negative and the other positive that are separated by a third unstable positive macrostate. Furthermore, assuming continuity  for the state variable, a Fokker-Planck equation was derived from the master equation of multistep processes. The extended absorbing boundary approach and a previous FPE based approach~\cite{Gardiner04} were used to calculate the MFPT of the transitioning of charge between stable macrostates in bistable charging of grains for various grain radii. The MFPTs calculated by two approaches for a given grain radius were in excellent agreement except for very small grains.  Very small grains posses small net elementary charges that the continuity assumption of charge, critical in the FPE description, may not be valid. For a given grain radius, the MFPT for a transition from the negative macrostate to the positive one was substantially larger than that for a transition in a reverse order. The  dimensionless MFPT for a transition from the positive stable macrostate to the negative macrostate showed little sensitivity to the grain radius. On the other hand, with the increase of the grain radius, it dropped first and then increased for the transition  from the negative to the positive macrostate.   The PDF of FPT, calculated by the extended absorbing boundary approach, was found substantially wider for a transition from the positive to negative macrostate, as compared to a transition from the negative to the positive macrostate.  Also, the derived FPE was further simplified through a linearization approximation about a stationary macrostate to  obtain a linear FPE \cite{Kampen07}.  Such an approximation is applicable about a macrostate if it is the only macrostate or if it is  sufficiently distant from the rest of macrostates in a multiple macrostate situation. Using the linear FPE, two equations for calculation of dissipation and growth times were provided.    When simplified to one-step processes, they were consistent with the dissipation and growth time equations  previously provided \cite{MR97} for grain charging through one step processes.

\acknowledgements

This work was in part supported by the National Science Foundation through award PHY-1414552.

\section*{Appendix}
\appendix

\section{Calculation of MFPT in the Fokker-Planck equation}
\label{sec:FirstPassageTime}
Here, the methodology given by \citet{Gardiner04} and \citet{Kampen07}  for the calculation of MFPT for a continuous stochastic process governed by a Fokker-Planck equation is presented. For a given process, this equation is available in two different forms, known as the forward equation and the backward equation. However, they  are equivalent, as discussed by \citet{Gardiner04}.  The forward Fokker-Planck equation, which is commonly referred just as the Fokker-Planck equation, reads
\begin{equation}
    \frac{\partial P(y,t)}{\partial t}=-\frac{\partial }{\partial y}A(y)P+\frac{1}{2}\frac{\partial^2}{\partial y^2}B(y)P. 
\label{eq:FPAppendix}
\end{equation}
\noindent
The mean passage time $\tau(y)$ associated with this equation, obeys 
\begin{equation}
   A(y)\frac{d \tau }{d y}+\frac{1}{2}B(y)\frac{d^2\tau}{d y^2}=-1, 
\label{eq:MFPT_FP}
\end{equation}
which is derived by the use of a backward equation equivalent to eq.~(\ref{eq:FPAppendix}), as illustrated by \citet{Gardiner04}. Eq.~(\ref{eq:MFPT_FP}) can be solved by direct integration. Three solutions have been previously developed for the range $L < y < R$, using three different sets of left and right boundary conditions at $y=L$ and $y=R$.  \citet{Gardiner04} provided the solution below when both boundary conditions are absorbing, i.e., $\tau(L)=\tau(R)=0$, is: 
\begin{eqnarray}
   \tau(y)&=&2\left[\int_L^y\frac{dy'}{\psi(y')}\int_y^R\frac{dy'}
   {\psi(y')}\int_L^{y'}\frac{dy''\psi(y'')}{B(y')}\right.\nonumber\\ 
   &&-\left.\int_y^R\frac{dy'}{\psi(y')}\int_L^y\frac{dy'}{\psi(y')}\int_L^{y'}\frac{dy''\psi(y'')}{B(y')} \right] \nonumber \\ 
&&   \left/{\int_L^R\frac{dy'}{\psi(y')}}\right., 
   \label{eq:leftOrRight}
\end{eqnarray}
\noindent
where 
\begin{equation}
    \psi(y)=\int_L^y\frac{2A(y')}{B(y')}dy'.
\end{equation}
\citet{Gardiner04} and \citet{Kampen07} gave the following solution to eq.~(\ref{eq:MFPT_FP}) when the right BC is absorbing $\tau(R)=0$ and the left BC is reflecting $d\tau/dy=0$ at $y=L$:

\begin{equation}
   \tau(y)=2\int_y^R \frac{dy'}{e^{\psi(y')}}\int_L^{y'}\frac{ e^{\psi(y'')}}{B(y'')}dy''.
\label{eq:tauRight}
\end{equation}
Similarly, when the left BC is absorbing, i.e.,  $\tau(L)=0$ and the right BC is reflecting, i.e., $d\tau/dy=0$ at $y=R$, the solution is \cite{Gardiner04}:

\begin{equation}
   \tau(y)=2\int_L^y \frac{dy'}{e^{\psi(y')}}\int_{y'}^R\frac{ e^{\psi(y'')}}{B(y'')}dy''.
   \label{eq:rightToLeft}
\end{equation}


\section{Calculation of the currents}
\label{sec:Currents}

The electron and ion currents to the the grain is calculated by the following equations in a Maxwellian plasma    \cite{Kimura1998,shotorban2014intrinsic,shotorban2015bistable}

\begin{equation}
I_e(Z)=  \Gamma\times \left\{ 
\begin{array}{l l}
  1+{Z\over \Omega} & \quad Z\ge 0,\\\\
\exp\left ({Z\over \Omega} \right)& \quad Z< 0,\\ \end{array} \right.
\label{eq:elecCurrent}
\end{equation}

\begin{equation}
I_i(Z)=  \Gamma\hat{n}_i \sqrt{\hat{T}_i\over \hat{m}_i}\times \left\{ 
\begin{array}{l l}
  1-{Z\over \hat{T}_i\Omega} & \quad Z\le 0,\\\\
\exp\left (-{Z\over \hat{T}_i\Omega} \right)& \quad Z> 0,\\ \end{array} \right.
\label{eq:ionCurrent}
\end{equation}

\noindent
where  $\hat{T}_i=T_i/T_e$, $\hat{m}_i=m_i/m_e$, and $\hat{n}_i=n_i/n_e$. The SEE current from the grain is calculated following Sternglass' theory \cite{sternglass1954theory,meyer1982flip,shotorban2015bistable} 

\begin{equation}
I_s(Z)=  3.7\delta_M\Gamma\times \left\{ 
\begin{array}{l l}
  \left(1+{Z\over{\Omega\hat{T}_s}}\right)\exp\left(-{Z\over{\Omega\hat{T}_s}}\right.
  \\ \left.+{Z\over\Omega}\right )F_{5,B}\left( {E_M\over{4k_BT_e}}\right)& \smallskip Z\ge 0,\\\\
\exp\left ({Z\over \Omega} \right)F_5\left( {E_M\over{4k_BT_e}}\right)& \smallskip Z< 0,\\ \end{array} \right.
\label{eq:SEE}
\end{equation}

\noindent 
where 
\[ F_5(x)=x^2\int_0^\infty  u^5\exp\left(-xu^2-u\right)du,\] 
\[F_{5,B}(x)=x^2\int_B^\infty  u^5\exp\left(-xu^2-u\right)du,\]
where $B=\sqrt{{4k_BT_e}Z/\Omega E_M}$  and $\hat{T}_s=T_s/T_e$ where $T_s$ is the temperature of the emitted secondary electrons.  For the definition of remaining parameters in this equation, readers are referred to  \S\ref{sec:stochasticCharging}.

%





\end{document}